 \documentclass[nofootinbib, prl, twocolumn, groupedaddress, showpacs, floatfix, preprintnumbers]{revtex4-1}
 \usepackage{amsmath, amstext, amssymb, amsfonts,amsmath}
 \usepackage{natbib,url,textcase,bm,color}
 \usepackage[dvips]{graphicx}
 \usepackage[dvips, colorlinks,urlcolor=blue,citecolor=blue,linkcolor=blue]{hyperref}
 \newcommand{\rd}{\mathrm{d}}
\begin{document}
%-----------------------------------------------------------------------
\title{The transverse-momenta distributions in high-energy $pp$ collisions\\ --- A statistical-mechanical approach}
%-----------------------------------------------------------------------
\author{
Leonardo J.\ L.\ Cirto$^{1}$, Constantino Tsallis$^{1,2}$,
Cheuk-Yin Wong$^{3}$ and Grzegorz Wilk$^{4}$ }
%-----------------------------------------------------------------------
\affiliation{$^1$Centro Brasileiro de Pesquisas Fisicas \&
National Institute of Science and Technology for Complex Systems, Rua Xavier Sigaud 150, 22290-180 Rio de Janeiro-RJ, Brazil}
\affiliation{$^2$Santa Fe Institute, 1399 Hyde Park Road, Santa Fe, NM 87501, USA}
\affiliation{$^3$Physics Division, Oak Ridge National Laboratory, Oak Ridge, Tennessee 37831, USA}
\affiliation{$^4$National Centre for Nuclear Research, Warsaw 00-681, Poland}
%-----------------------------------------------------------------------
\begin{abstract}
We analyze  LHC available data measuring the distribution
probability of transverse momenta~$p_T$ in proton-proton
collisions at $\sqrt{s}=0.9\,\textrm{TeV}$ (CMS, ALICE, ATLAS) and
$\sqrt{s}=7\,\textrm{TeV}$ (CMS, ATLAS).  A remarkably good
fitting can be obtained, along fourteen decades in magnitude, by
phenomenologically using $q$-statistics for a {\it single} particle of a
two-dimensional relativistic ideal gas.
  The parameters that have been obtained by assuming $\rd N/ p_T \rd p_T \rd y \propto e_q^{-E_T/T}$ at mid-rapidity  are,  in all cases, $q \simeq 1.1$ and $T\simeq 0.13\,\textrm{GeV}$  (which satisfactorily compares with the pion mass).
  This fact suggests the approximate validity of a  ``no-hair" statistical-mechanical description of the hard-scattering hadron-production process in which the detailed mechanisms of parton scattering, parton cascades, parton fragmentation, running coupling and other information can be subsumed under the stochastic dynamics in the lowest-order description.
In addition to that basic structure, a finer analysis of the data suggests a small
oscillatory structure on top of the leading $q$-exponential.
The physical origin of such intriguing oscillatory behavior remains
elusive, though it could be related to some sort of fractality or scale-invariance within the system.
\end{abstract}
%-----------------------------------------------------------------------
 \maketitle
%-----------------------------------------------------------------------
The transverse momentum distributions of hadrons
produced in collisions involving protons and also heavy ions at
RHIC and LHC energies~\cite{LISTA1} have recently been successfully
investigated through nonextensive statistical
mechanics~\cite{LISTA2}. A statistical-mechanical approach to
analytically describe these quantities was successfully used
phenomenologically by Hagedorn four decades ago~\cite{Hagedorn}.
By adopting the idea of some kind of thermal quasi-equilibrium, he
introduced the Boltzmann-Gibbs (BG) factor within his theoretical
model. The Hagedorn proposal works very well at low beam-energies.
However, the experiments at higher energies, which have been
achieved with the modern experiments at LHC and other colliders,
have produced data that neatly depart more and more from a
standard thermal behavior, giving rise to power-law distributions
in the asymptotic high-$p_T$ regime.

In order to explain the discrepancy between the BG distribution and experiments at high~$p_T$, a number of works have been done through changing the~BG thermostatistical start-point by its nonextensive generalization.
Broadly speaking, this change corresponds to replacing the BG
exponential weight by  its nonextensive $q$-ex\-po\-nen\-tial  deformation.
Hagedorn himself made heuristically a somewhat similar approach.
With the aim of extending his analysis to higher~$p_T$ values,
power-law functions conforming to pQCD~\cite{Ten12} were used instead of
exponentials, without resorting to thermostatistical
justifications~\cite{Hagedorn, Arnison_UA1_Collaboration}\footnote{Actually, such approach
was firstly proposed  in the analysis of the uncorrelated jet model,
see~\cite{Michael}.}.
Theories  proposed within the nonextensive
thermostatistics scenario include, for example, the generalized
version of the Hagedorn asymptotic bootstrap
principle~\cite{BediagaCuradoMiranda_PA2000, Deppman_PA2012MarquesAndradeDeppmanPRD2013, Beck_PA2000},
%$q$-entropy~\cite{Bir13} ??,
reservoir fluctuations~\cite{Bir14},
micro-canonical jet-fragmentation~\cite{Urm12},
consequences of thermodynamic consistency~\cite{Clw}, and a relativistic
hard-scattering model in perturbative QCD~\cite{WongWilk_APPB2012, WongWilk_PRD2013}.
This approach gets also its impact in description of nuclear matter
under extreme conditions (as those encountered in present heavy ion collisions), cf.,
for example,~\cite{qNuclear} and references therein.

Recently, the experimental data reveal surprisingly that the
transverse momentum spectrum from the very low energy regime of
many tenths of a GeV to the very high energy regime of 
hundreds of GeV, in high-energy~$pp$ collisions at central rapidities, is
characterized by a small number of degrees of
freedom~\cite{WongWilk_APPB2012}. However, one expects on physical
grounds at least three different physical mechanisms contributing
to the hadron production process. In the low~$p_T$ region at
central rapidities, the mechanism of  non-perturbative
\emph{string-fragmentation} is expected to play an important role~\cite{And83, Won94}.
In this mechanism, the hadrons are produced
from the string stretched between the colliding nucleons with a
plateau structure in rapidity. In the low~$p_T$ region near the beam
and target rapidities, the mechanism of
\emph{direct-fragmentation} is expected to play a dominant role~\cite{Won80}.
In this mechanism, the produced hadron fragments
directly from the nucleon without making a collision with the
partons of the other colliding nucleon. The transverse spectrum
then depends on the probability of the hadron fragmenting out of the nucleon.
In the high~$p_T$ region, the \emph{relativistic hard-scattering process}
becomes important. In this mechanism, a
parton from the projectile nucleon scatters elastically with a
parton from the target nucleon, and the parton subsequently cascades and  fragments  into the produced
hadrons. The transverse spectrum therefore depends on many factors,
including the parton distribution in the nucleons, the
parton-parton hard-scattering amplitude, the cascade and  fragmentation of the
scattered parton to the hadrons, as well as the running of the
coupling constant, as described in~\cite{WongWilk_PRD2013}.

%The string fragmentation, the direct-fragmentation and the hard-scattering processes
The three mechanisms mentioned above (string- and direct-fragmentations, and hard-scattering) are expected to give rise to
different shapes of the transverse distributions as they depend on the
transverse momentum in different ways. Indeed, the string
fragmentation is associated with a flux tube and the transverse
momentum distribution is limited by the flux tube dimension.
The direct fragmentation is governed by the parton intrinsic~$p_T$
motion inside a nucleon and the gluon radiations, whereas
 the transverse momentum distribution in hard-scattering is governed by the law of parton-parton scattering, parton distribution, parton cascade, parton fragmentation, and the running of the coupling constant.
Therefore, one would normally expect that the three different
production mechanisms will lead to different behaviors as
functions of~$p_T$, and there would consistently be a breakdown of
a single description at some point of the complete
spectrum.

What interestingly emerges from the analysis of the data in
high-energy~$pp$ collisions is that the good agreement of the present phenomenological fit extends to the {\it whole}~$p_T$
region (or at least for~$p_T$ greater than $0.5\,\textrm{GeV}/c$, where reliable experimental data are available; in fact, the ALICE $0.9 \, \textrm{TeV}$ data reliably extend even to lower values, namely down to $p_T \simeq 0.1 \,\textrm{GeV}/c$)~\cite{WongWilk_APPB2012}.
This being achieved with very few free parameters implies the simplicity of the underlying structure
and the dominance of one of the three mechanisms
over virtually the whole~$p_T$ range
in the central rapidity region. It is reasonable to
consider the dominant mechanism to be the hard-scattering process
because the other two mechanisms are unlikely to produce hadrons
with high~$p_T$. Moreover, the dominance of hard-scattering also
for the production of low-$p_T$  hadron in the central rapidity region is supported by
two-particle correlation data where  the two-body correlations in
minimum $p_T$-biased data reveal that a produced hadron is
correlated  with a ``ridge" of particles along a wide range
of~$\Delta\eta$ on the azimuthally away side centering around
$\Delta\phi\sim \pi$~\cite{STAR06,Abe12,Ray11}.
The $\Delta\phi\sim \pi$ (back-to-back) correlation indicates that the correlated pair is
related by a collision, and the $\Delta\eta$  correlation in the
shape of a ridge indicates that  the two particles are partons
from the two nucleons and they carry fractions of the longitudinal
momenta of their parents, leading to the ridge of~$\Delta \eta$ at
$\Delta\phi\sim \pi$.

While the basic hard-scattering process is relatively simple in the jet
production level, there are many layers of stochastic processes in
the production of hadrons from jets which mask this simplicity.
The hard-scattering process exhibits the power-law behavior of the
transverse momentum in its basic framework for jet
production~\cite{WongWilk_APPB2012}. However, many other stochastic elements 
are involved in this process which definitively increase its
complexity.

In spite of this fact, it turns out, remarkably enough, to be reasonable
%However from jet to hadrons, there are many elements of stochastic processes in the sequential parton cascade,
%and parton fragmentation,  final-state interactions, and the running of the coupling constant. Even though the relativistic
%hard-scattering processes yields a simple power law, those other common elements involving stochastic ergodic dynamics with random elements may play a significant role in modifying the power law.
%Under the circumstance of such domination of complex ergodic dynamics, it becomes reasonable
to assume,  in the lowest-order description, a ``no-hair"
statistical-mechanical hypothesis for the 
% whole
transverse-momentum distribution in the~$p_T$ region above $0.5\,\textrm{GeV}/c$.
The hadron production process appears to be well
characterized by a single-particle statistical-mechanical
description of QCD quanta in an relativistic $d=2$ system.
All other information about the produced hadron matter, such as
the parton distribution, the hard-scattering mechanism  of jet
production, jet-parton cascade process followed by parton fragmentation, and
the intermediary possible quark-gluon plasma, ``disappear" behind
the stochastic process. 
%Let us emphasize, at this point, that the effective density of states to be adopted should be the generalization of the 
%Boltzmann-Gibbs description for a $d$-dimensional  single particle.

Towards such a goal of description, we can postulate that for the
whole range of~$p_T$ from about $0.5\,\textrm{GeV}/c$ to very high values, the
transverse distribution of the produced hadrons obeys the
``no-hair'' statistical-mechanical description with very few
degrees of freedom of the hard-scattering process. Motivated by
the good agreement of previous works, we use here the nonextensive
thermostatistical background to examine the experimental
high-energy~$pp$ data and we describe the distribution of hadronic
transverse momenta at central rapidity $y\simeq 0$  with the following
nonextensive ansatz\footnote{We are adopting unity for the Boltzmann
constant~$k_B$.}
\begin{equation}
% \frac{dN}{dy d\boldsymbol{p}_T} = \frac{1}{2\pi p_T} \frac{dN}{dy dp_T} = A e_q^{-E_T/T} \,, 
\frac{\rd N}{\rd y \rd\boldsymbol{p}_T} = \frac{1}{2\pi p_T} \frac{\rd N}{\rd y \rd p_T} = A e_q^{-E_T/T} \,, 
\label{qexponential}
\end{equation}
where the $q$-exponential function $e_q^z$ is defined by
\begin{equation}
e_q^z \equiv \left[ 1 + \left(1-q\right)z \right]^{1/(1-q)}
\;\;\;(e_1^z=e^z).
\end{equation}
Here, $E_T$ is the relativistic energy associated with the
transverse momentum of a single particle of the beam, namely, $E_T
= \sqrt{\boldsymbol{p}_T^2 + m^2}$. As most of the produced
particles are pions, $m$ was taken equal to the meson
mass~$m_{\pi}$. Notice that, if the distributions are properly
normalized, the constant prefactor~$A$ is {\it not} an independent
parameter but a straightforward function of $(q,T)$.
% mass~$m_{\pi}=0.135\,\textrm{GeV}/c^2$.

\begin{figure}%[b] %\vspace{-0.3cm}
\centering
\includegraphics[width=1.00\linewidth]{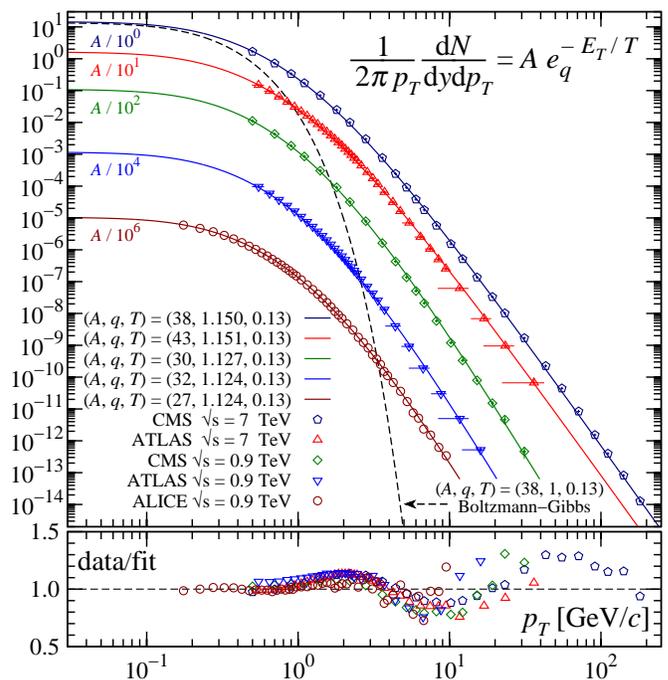}
\caption{Comparison of Eq.~\eqref{qexponential} with the
experimental transverse momentum distribution of hadrons in~$pp$
collisions at central rapidity $y$. Herein the temperature is
being measured in $[T]=\textrm{GeV}$ and the normalization
constant in $[A]=\textrm{GeV}^{-2}/c^3$.
For a better visualization both the data and the analytical curves have been
divided by a constant factor as indicated.
The ratios data/fit are shown at the bottom, where a nearly log-periodic behavior is
observed on top of the $q$-exponential one.
Data from~\cite{WongWilk_APPB2012,WongWilk_PRD2013}.
}
\label{figure}
\end{figure}

The experimental results for two energies ($0.9$ and $7\,\textrm{TeV}$) and
various detectors (CMS, ALICE,
ATLAS)~\cite{WongWilk_APPB2012,WongWilk_PRD2013} together with the
proposed distribution~\eqref{qexponential} are indicated in
Fig.~\ref{figure}. We verify that~$q$  increases slightly with the
beam energy, but, for the present energies, remains always $q\simeq 1.1$.
Some indications exist that, in the limit of
extremely high energies, $q$ approaches a limiting value (possibly
close to 1.2~\cite{Beck_PA2000, Wibig_JPG2010}).
The effective temperature is, in all cases, $T\simeq 0.13\,\textrm{GeV}$, neatly compatible with the meson mass 
$m_{\pi} = 0.135\,\textrm{GeV}/c^2$. The dashed line
(an ordinary exponential of~$E_T$) illustrates how important can be the discrepancy with
the analogous Boltzmann-Gibbs approach as~$p_T$ increases.

What we may extract from the behavior of the experimental data is
that the Hagedorn scenario appears to be essentially correct
excepting for the fact that we are \emph{not} facing thermal
equilibrium but a different type of stationary state, typical of
violation of ergodicity (for a discussion of the kinetic and
effective temperatures
see~\cite{Effective_Temperature,overdamped}). These results
reinforce the pioneering connection between quantum chromodynamics
and nonextensive statistics first suggested by Walton and
Rafelski~\cite{WaltonRafelski2000}. The successful fitting of
Eq.~\eqref{qexponential} along impressive 14 decades strongly
points at the present ``no-hair" assumption
(namely a two-dimensional relativistic single-particle system)
as the simplest one on which further sophisticated models can be built.
We emphasize also that, {\it in all cases}, the
temperature turns out to be one and the same, namely $T=0.13\,\textrm{GeV}$.

As a concluding remark,  we note that the data/fit plot in the
bottom part of Fig.~\ref{figure} exhibits intriguing (rough)
log-periodic oscillations, which suggest some hierarchical
fine-structure in the quark-gluon system where hadrons are
generated. This behavior is possibly an indication of some kind of
(multi)fractality in the system. Indeed, the concept of
\emph{self-similarity}, one of the landmarks of fractal
structures, has been used by Hagedorn in his very definition of
fireball, as was previously pointed out by Beck~\cite{Beck_PA2000}
and has been found in the analysis of jets produced in~$pp$ collisions at LHC~\cite{GWZW}.
These small oscillations have already been preliminary
discussed in~\cite{Wilk1,Wilk2}, where the authors were able to
mathematically accommodate them by essentially
allowing the index~$q$ in Eq.~\eqref{qexponential} to be a complex
number\footnote{It should be
noted here that an alternative to complex~$q$ would be a log-periodic
fluctuating scale parameter~$T$: such possibility was discussed in~\cite{Wilk2}.}
(see also Refs.~\cite{logperiodic,Sornette1998}).

%-----------------------------------------------------------------------
\begin{acknowledgments}
Two of us (L.J.L.C. and C.T.) have benefited from partial
financial support from CNPq, Faperj and Capes (Brazilian
agencies). The research of CYW was supported in part by the
Division of Nuclear Physics, U.S. Department of Energy, and that of GW
was supported by the National Science Center (NCN) under contract
DEC-2013/09/B/ST2/02897.
\end{acknowledgments}
%-----------------------------------------------------------------------
% BIBLIOGRAPHY:
%-----------------------------------------------------------------------
% \clearpage
  
%-----------------------------------------------------------------------

\begin{thebibliography}{99}
%-----------------------------------------------------------------------
% HIGH ENERGY:
%-----------------------------------------------------------------------
\bibitem{LISTA1}
V.\ Khachatryan \emph{et al}.\ (CMS Collaboration), 
Phys.\ Rev.\ Lett.\ \textbf{105}, 022002 (2010);
J.\ High Energy Phys.\ \textbf{02}, 041 (2010);
%
S.\ Chatrchyan \emph{et al}.\ (CMS Collaboration)
% J.\ High Energy Phys.\ \textbf{08}, 086 (2011);
\emph{ibid}.\ \textbf{08}, 086 (2011);
%
G.\ Aad \emph{et al}.\ (ATLAS Collaboration),
New J.\ Phys.\ \textbf{13}, 053033 (2011);
%
K.\ Aamodt \emph{et al}.\ (ALICE Collaboration),
Phys.\ Lett.\ B    \textbf{693}, 53 (2010);
Eur.\ Phys.\ J.\ C \textbf{71}, 1594 (2011); \textbf{71} 1655 (2011);
%
A.\ Adare \emph{et al}.\ (PHENIX Collaboration),
Phys.\ Rev.\ D \textbf{83}, 052004 (2011);
Phys.\ Rev.\ C \textbf{83}, 064903 (2011);
%
B.\ Abelev \emph{et al}.\ (ALICE Collaboration),
Phys.\ Lett.\ B \textbf{722}, 262 (2013).

%-----------------------------------------------------------------------
% q-STATISTICS:
%-----------------------------------------------------------------------
\bibitem{LISTA2}
C. Tsallis,
J.\ Stat.\ Phys.\ \textbf{52}, 479 (1988);
%
% C. Tsallis, 
\emph{Introduction to Nonextensive Statistical Mechanics -- Approaching A Complex World} (Springer, New York, 2009);
\emph{Entropy}, in {\it Encyclopedia of Complexity and Systems Science}, edited by R.\ A.\ Meyers (Springer, Berlin, 2009) p.\ 2859;
%
M.\ Gell-Mann and C.\ Tsallis eds.,
\emph{Nonextensive Entropy -- Interdisciplinary Applications} (Oxford University Press, New York, 2004).
%
A regularly updated bibliography on nonadditive entropies and nonextensive
statistical mechanics is available at \url{http://tsallis.cat.cbpf.br/biblio.htm}.
%-----------------------------------------------------------------------

\bibitem{Hagedorn}
R.\ Hagedorn,
%Multiplicities, {\it $p_T$ distributions and the expected hadron$\rightarrow$quark-gluon phase transition},
Rev.\ Nuovo Cimento \textbf{6}, 1 (1983).
% url={http://dx.doi.org/10.1007/BF02740917}
% doi={10.1007/BF02740917}

\bibitem{Ten12}
R.\ Blankenbecler and S.\ J.\ Brodsky,
Phys.\ Rev.\ D {\bf 10}, 2973 (1974);
A.\ L.\ S.\ Angelis \emph{et~ al}., 
Phys.\ Lett.\ B {\bf 79}, 505 (1978);
for the history of the power law, see
J.\ Rak and M.\ J.\ Tannenbaum, 
{\it High-$p_T$ Physics in the Heavy Ion Era} (Cambridge University Press, Cambridge, 2013).

\bibitem{Arnison_UA1_Collaboration}
G.\ Arnison \emph{et al} (UA1 Collaboration),
%{\it Transverse momentum spectra for charged particles at the CERN proton-antiproton collider},
Phys.\ Lett.\ B \textbf{118}, 167 (1982).
%doi="http://dx.doi.org/10.1016/0370-2693(82)90623-2",
%url="http://www.sciencedirect.com/science/article/pii/0370269382906232",

\bibitem{Michael}
C.\ Michael and L.\ Vanryckeghem,
J.\ Phys.\ G \textbf{3}, L151 (1977);
C.\ Michael, 
Prog.\ Part.\ Nucl.\ Phys.\ \textbf{2}, 1 (1979).

\bibitem{BediagaCuradoMiranda_PA2000}
I.\ Bediaga, E.\ M.\ F.\ Curado,  and J.\ M.\ de Miranda,
%{\it A nonextensive thermodynamical equilibrium approach in $e+e\rightarrow$ hadrons},
Physica A \textbf{286}, 156 (2000).
% doi = "http://dx.doi.org/10.1016/S0378-4371(00)00368-X",
% url = "http://www.sciencedirect.com/science/article/pii/S037843710000368X",

%-----------------------------------------------------------------------
% \bibitem{Deppman_PA2012}
% A.\ Deppman,
% %{\it Self-consistency in non-extensive thermodynamics of highly excited hadronic states},
% Physica A \textbf{391}, 6380 (2012).
% % doi = "http://dx.doi.org/10.1016/j.physa.2012.07.071",
% % url = "http://www.sciencedirect.com/science/article/pii/S0378437112007625",
%
% \bibitem{MarquesAndradeDeppmanPRD2013}
% L.\ Marques,  E.\ Andrade-II, and A.\ Deppman,
% % {\it Nonextensivity of hadronic systems},
% Phys.\ Rev.\ D \textbf{87}, 114022 (2013).
% % doi = {10.1103/PhysRevD.87.114022},
% % url = {http://link.aps.org/doi/10.1103/PhysRevD.87.114022}
%
% \bibitem{DeppmanJPG2014}
% A.\ Deppman,
% % Properties of hadronic systems according to the nonextensive self-consistent thermodynamics
% J.\ Phys.\ G.\ \textbf{41}, 055108 (2014).
% % url={http://stacks.iop.org/0954-3899/41/i=5/a=055108},
% % doi:10.1088/0954-3899/41/5/055108

\bibitem{Deppman_PA2012MarquesAndradeDeppmanPRD2013}
A.\ Deppman,
Physica A \textbf{391}, 6380 (2012);
% A.\ Deppman,
J.\ Phys.\ G.\ \textbf{41}, 055108 (2014);
L.\ Marques,  E.\ Andrade-II, and A.\ Deppman,
Phys.\ Rev.\ D \textbf{87}, 114022 (2013).
%-----------------------------------------------------------------------

\bibitem{Beck_PA2000}
C.\ Beck,
%{\it Non-extensive statistical mechanics and particle spectra in elementary interactions},
Physica A \textbf{286}, 164 (2000).
% doi = "http://dx.doi.org/10.1016/S0378-4371(00)00354-X",
% url = "http://www.sciencedirect.com/science/article/pii/S037843710000354X",

%\bibitem{Bir13}T.\ S.\ Bir\'o,Physica A {\bf 392}, 3132 (2013).
% Ideal gas provides qq-entropy
% http://dx.doi.org/10.1016/j.physa.2013.03.028

\bibitem{Bir14}
T.\ S.\ Bir\'o, G.\ G.\ Barnaf\"oldi, P.\ V\'an, and K.\ \"Urm\"ossy,
arXiv:1404.1256 [hep-ph].
% Statistical Power-Law Spectra due to Reservoir Fluctuations
% http://arxiv.org/abs/1404.1256

\bibitem{Urm12}
K.\ \"Urm\"osy, G.\ G.\ Barnaf\"oldi, and T.\ S.\ Bir\'o,
Phys.\ Lett.\ B {\bf 718}, 125 (2012);
T.\ S.\ Bir\'o, K.\ \"Urm\"osy, and Z.\ Schram,
J.\ Phys.\ G {\bf 37}, 094027 (2010);
T.\ S.\ Bir\'o and P.\ V\'an,
Phys.\ Rev.\ E {\bf 83}, 061147 (2011);
T.\ S.\ Bir\'o and Z.\ Schram,
EPJ web conf.\ {\bf 13}, 05004 (2011);
T.\ S.\ Bir\'o,
% {\it Is there a temperature? Conceptual Challenges at High Energy, Acceleration and Complexity}
{\it Is there a temperature?} (Springer, New York, 2011).

\bibitem{Clw} 
J.\ Cleymans and D.\ Worku,
Eur.\ Phys.\ J.\ A {\bf 48}, 160 (2012);
% http://dx.doi.org/10.1140/epja/i2012-12160-0
J.\ Phys.\ G {\bf 39}, 025006 (2012);
% http://dx.doi.org/10.1088/0954-3899/39/2/025006
M.\ D.\ Azmi and J.\ Cleymans,
% J.\ Phys.\ G.\ \textbf{41} 065001 (2014).
\emph{ibid}.\ \textbf{41}, 065001 (2014).
%url={http://stacks.iop.org/0954-3899/41/i=6/a=065001},


\bibitem{WongWilk_APPB2012}
C.\ Y.\ Wong and G.\ Wilk,
%{\it Tsallis Fits to $p_T$ Spectra for $pp$ Collisions at LHC},
Acta Phys.\ Polonica B \textbf{43}, 2043  (2012).

\bibitem{WongWilk_PRD2013}
C.\ Y.\ Wong and G.\ Wilk,
%{\it Tsallis fits to ${p}_{T}$ spectra and multiple hard scattering in $pp$ collisions at the LHC},
Phys.\ Rev.\ D \textbf{87}, 114007 (2013).
% doi = {10.1103/PhysRevD.87.114007},
% url = {http://link.aps.org/doi/10.1103/PhysRevD.87.114007}

\bibitem{qNuclear}
A.\ P.\ Santos, F.\ I.\ M.\ Pereira, R.\ Silva, and J.\ S.\ Alcaniz, 
J.\ Phys.\ G \textbf{41}, 055105 (2014);
J.\ Ro\.zynek and G.\ Wilk, 
% J.\ Phys.\ G \textbf{36}, 125108 (2009);
\emph{ibid}.\ \textbf{36}, 125108 (2009);
% http://dx.doi.org/10.1088/0954-3899/36/12/125108
EPJ web conf.\ \textbf{13}, 05002 (2011).
% http://dx.doi.org/10.1051/epjconf/20111305002


\bibitem{And83} 
B.\ Andersson, G.\ Gustafson, and T.\ Sj\"ostrand, 
% Zeit.\ f{\"u}r Phys.\ C {\bf 20}, 317 (1983); 
Z.\ Phys.\ C {\bf 20}, 317 (1983); 
B.\ Andersson, G.\ Gustafson, G.\ Ingelman, and T.\  Sj\"ostrand, 
Phys.\  Rep.\ {\bf 97}, 31 (1983);
T.\ Sj\"ostrand and M.\ Bengtsson, 
Comput.\ Phys.\ Commun.\  {\bf 43}, 367 (1987);
B.\ Andersson, G.\ Gustavson, and B. Nilsson-Alqvist,
Nucl.\ Phys.\ B {\bf 281}, 289 (1987).

\bibitem{Won94}
C.\ Y.\ Wong,
{\it Introduction to High-Energy Heavy-Ion Collisions}
(World Scientific Publishing Company, Singapore, 1994).

\bibitem{Won80}
C.\ Y.\ Wong and R.\ Blankenbecler,
%\emph{Direct fragmentation and hard-scattering processes in relativistic heavy-ion reactions},
Phys.\ Rev.\ C {\bf 22}, 2433 (1980).
% doi = {10.1103/PhysRevC.22.2433},


\bibitem{STAR06}
J.\ Adams \emph{et~al}.\ (STAR Collaboratotion),
Phys.\ Rev.\ D \textbf{74}, 032006 (2006).


\bibitem{Abe12}
B.\ Abelev \emph{et~al}.\ (ALICE Collaboration),
Phys.\ Rev.\ D {\bf 86}, 112007 (2012).

\bibitem{Ray11}
R.\ L.\ Ray, 
Phys.\ Rev.\ D {\bf 84}, 034020 (2011);
T.\ A.\ Trainor and D.\ J.\ Prindle,
arXiv:1310.0408 [hep-ph].
% Improved isolation of the p-p underlying event based on minimum-bias trigger-associated hadron correlations

\bibitem{Wibig_JPG2010}
T.\ Wibig,
J.\ Phys. G: Nucl.\ Part.\ Phys.\ \textbf{37}, 115009 (2010).
% The non-extensivity parameter of a thermodynamical model of hadronic interactions at LHC energies 
% http://dx.doi.org/10.1088/0954-3899/37/11/115009

%-----------------------------------------------------------------------
% Effective Temperature:
%-----------------------------------------------------------------------
\bibitem{Effective_Temperature}
W.\ Niedenzu, T.\ Grie{\ss}er, and H.\ Ritsch,
%{\it Kinetic theory of cavity cooling and self-organisation of a cold gas},
Europhys.\ Lett.\ \textbf{96}, 43001 (2011);
%
L.\ A.\ Gougam and M.\ Tribeche,
%{\it Debye shielding in a nonextensive plasma},
Phys.\ Plasmas \textbf{18}, 062102 (2011);
%
L.\ A.\ Rios, R.\ M.\ O.\ Galv\~ao, and L.\ Cirto,
%{\it Comment on ``{Debye} shielding in a nonextensive plasma'' [Phys. Plasmas 18, 062102 (2011)]},
% Phys.\ Plasmas \textbf{19}, 034701 (2012);
\emph{ibid}.\ \textbf{19}, 034701 (2012);
%
L.\ J.\ L.\ Cirto, V.\ R.\ V.\ Assis, and C.\ Tsallis,
%{\it Influence of the interaction range on the thermostatistics of a classical many-body system},
Physica A {\bf 393}, 286 (2014).
%-----------------------------------------------------------------------

%-----------------------------------------------------------------------
% Overdamped:
%-----------------------------------------------------------------------
\bibitem{overdamped}
J.\ S.\ Andrade Jr., G.\ F.\ T.\ da Silva, A.\ A.\ Moreira, F.\ D.\ Nobre, and E.\ M.\ F.\ Curado,
%{\it Thermostatistics of overdamped motion of interacting particles},
Phys.\ Rev.\ Lett.\ {\bf 105}, 260601 (2010);
%
M.\ S.\ Ribeiro, F.\ D.\ Nobre and E.\ M.\ F.\ Curado,
%{\it Overdamped motion of interacting particles in general confining potentials: Time-dependent and stationary-state analyses},
Eur.\ Phys.\ J.\ B {\bf 85}, 399 (2012);
%
% M.\ S.\ Ribeiro, F.\ D.\ Nobre, and E.\ M.\ F.\ Curado,
%{\it Time evolution of interacting vortices under overdamped motion},
Phys.\ Rev.\ E {\bf 85}, 021146 (2012);
%
E.\ M.\ F.\ Curado, A.\ M.\ C.\ Souza, F.\ D.\ Nobre, and R.\ F.\ S.\ Andrade,
%{\it Carnot cycle for interacting particles in the absence of thermal noise},
% Phys.\ Rev.\ E  {\bf 89}, 022117 (2014).
\emph{ibid}.\ {\bf 89}, 022117 (2014).
%-----------------------------------------------------------------------

\bibitem{WaltonRafelski2000}
D.\ B.\ Walton and J.\ Rafelski,
%{\it Equilibrium distribution of heavy quarks in Fokker-Planck dynamics},
Phys.\ Rev.\ Lett.\ {\bf 84}, 31 (2000).

\bibitem{GWZW}
G.\ Wilk and Z.\ W{\l}odarczyk,
%{\it Self-similarity in jet events following from pp collisions at LHC},
Phys.\ Lett.\ B \textbf{727}, 163 (2013).

%-----------------------------------------------------------------------
% Log-Periodic:
%-----------------------------------------------------------------------
\bibitem{Wilk1}
G.\ Wilk and Z.\ W{\l}odarczyk,
%{\it Tsallis distribution with complex nonextensivity parameter q},
Physica A \textbf{413}, 53 (2014).

\bibitem{Wilk2}
G.\ Wilk and Z.\ W{\l}odarczyk,
% {\it Log-periodic oscillations of transverse momentum distributions},
ArXiv:1403.3508 [hep-ph].

\bibitem{logperiodic}
C.\ Tsallis, L.\ R.\ da Silva, R.\ S.\ Mendes, R.\ O.\ Vallejos, and A.\ M.\ Mariz,
%{\it Specific heat anomalies associated with Cantor-set energy spectra},
Phys.\ Rev.\ E {\bf 56}, R4922 (1997);
%
L.\ R.\ da Silva, R.\ O.\ Vallejos, C.\ Tsallis, R.\ S.\ Mendes, and S. Roux,
%{\it Specific heat of multifractal energy spectra},
% Phys.\ Rev.\ E {\bf 64}, 011104 (2001).
\emph{ibid}.\ {\bf 64}, 011104 (2001).
%-----------------------------------------------------------------------

\bibitem{Sornette1998}
D.\ Sornette,
%{\it Discrete-scale invariance and complex dimensions},
Phys.\ Rep.\ {\bf 297}, 239 (1998).

%\bibitem{D2}
% Since we are supposed to speak to particle physics community a few words plus some reference(s) explayning what we understand by relativistic d=2 system is necessary here; I'm not able to fill this point (GW).
%-----------------------------------------------------------------------
\end{thebibliography}
\end{document}